\documentclass[aps,prl,floatfix,twocolumn,superscriptaddress,showpacs,reprint]{revtex4}
\usepackage{graphicx}
\usepackage{mathrsfs}
\usepackage{amsmath}
\usepackage{subfigure}
\usepackage{bm}
\usepackage{verbatim}
\usepackage{color}
\usepackage{xcolor}
\usepackage[colorlinks=true, letterpaper=true, pdfstartview=FitV, linkcolor=blue, citecolor=blue, urlcolor=blue]{hyperref}
\usepackage{siunitx}

\begin{document}
\title{Non-Local Transport Mediated by Spin-Supercurrents}
\author{Hua Chen}
\affiliation{Department of Physics, The University of Texas at Austin, Austin, Texas 78712, USA}
\author{Andrew D. Kent}
\affiliation{Department of Physics, New York University, New York, New York 10003, USA} 
\author{Allan H. MacDonald} 
\affiliation{Department of Physics, The University of Texas at Austin, Austin, Texas 78712, USA}
\author{Inti Sodemann}
\affiliation{Department of Physics, The University of Texas at Austin, Austin, Texas 78712, USA}

\begin{abstract}
In thin film ferromagnets with perfect easy-plane anisotropy, the component of total spin perpendicular to the easy plane is 
a good quantum number and the corresponding spin supercurrent can flow without dissipation.  In this Letter we explain how 
spin supercurrents couple spatially remote spin-mixing vertical transport channels, even when easy-plane anisotropy is 
imperfect, and discuss the possibility that this effect can be used to fabricate new types of electronic devices.   
\end{abstract}
\pacs{75.76.+j, 75.70.-i, 72.25.Pn, 85.75.-d}
\maketitle

\noindent
{\color{blue}{\em Introduction}}---
Recent progress~\cite{Brataas2012} in the growth of magnetic thin films has made it possible to 
construct circuits in which materials with perpendicular and 
in-plane magnetic anisotropy are flexibly combined.
As we explain below these advances have improved prospects for the experimental realization of a new 
class of effects in spintronics in which collective magnetic degrees of freedom play a more active role.
In this Letter we discuss effects which rely on 
the ability of ferromagnetic materials with strong easy-plane order to carry
dissipationless spin supercurrents~\cite{Sonin,Konig,jj,Tserkovnyak,takei2014} that are in many ways analogous to the dissipationless 
charge currents carried by superconductors. In both XY ferromagnets and superconductors 
the energy of the ordered state is independent of an angle $\phi$, 
the azimuthal magnetic orientation angle in the XY ferromagnet case and the 
Cooper-pair condensate phase angle in the superconducting case.  
States with a definite value of this angle break spin-rotational symmetry around the $\hat{z}$ axis 
and gauge symmetry respectively.  The analogy between XY ferromagnetism and superconductivity extends to superflow behavior.
Just as superconductors can support dissipationless charge currents, perfect easy-plane 
ferromagnets can support dissipationless currents~\cite{Sonin,Konig,jj,Tserkovnyak,takei2014} of the conserved $\hat{z}$ 
component of total spin.  

\begin{figure}[!]
\includegraphics[scale=0.35]{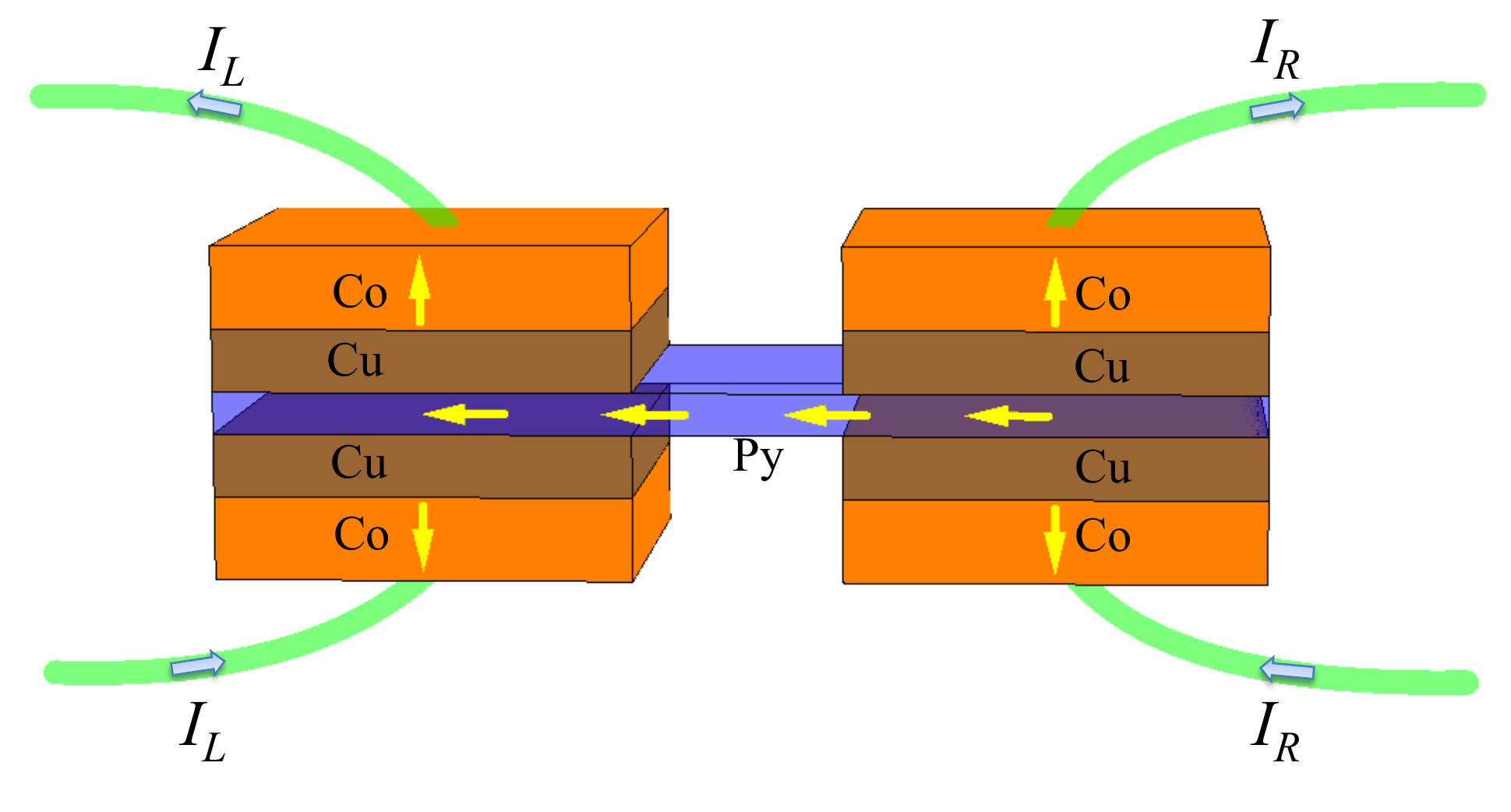}
\caption{\label{Schematic}(color online). 
Charge transport through two (Left and Right) metallic multilayer stacks containing both
perpendicular-anisotropy films and a common easy-plane anisotropy thin film. Transport electrons in both metal 
stacks exert spin-transfer torques on the easy-plane anisotropy film.  
Exchange interactions within that film couple the two vertical transport channels. When the orientation of the easy plane 
ferromagnet is time-independent the left (L) and right (R) conductances are independent.  When the total current 
$I_{L}+I_{R}$ exceeds a critical value, spin-transfer torques drive the easy-plane thin-film nano-magnet into a dynamic precessional 
limit cycle in which the transport properties of the L and R stacks are interdependent. If the Gilbert-damping coefficient of the 
easy-plane magnet is small, the {\it dc} conductance of both stacks is strongly suppressed in the dynamic configuration.
The specific materials indicated in this illustration are discussed in the text.  
}
\end{figure}

As an illustration of the types of effects that can occur we address 
specifically the super-spintronic circuit proposed in Ref.~\onlinecite{Konig}
and illustrated in Fig.~\ref{Schematic}. The circuit consists of two thin film metal stacks 
containing perpendicular anisotropy magnetic layers and 
linked by a common easy-plane ferromagnetic thin film, assumed here to be permalloy.  
As we will describe, collective magnetic behavior induced by exchange interactions within the 
easy-plane material lead to non-local effects that can be  
much stronger than the familiar effects associated with spin-diffusion in normal metals.  
The cross-sectional area of the two stacks and the length of the 
permalloy link between them are assumed to be sufficiently large to avoid 
significant charge transport cross-talk~\cite{Caveat_CrossTalk,Janak,Bass} between stacks.
The magnetization of the perpendicularly magnetized metal stacks is assumed to be pinned and only the easy-plane thin film ferromagnet's magnetization is allowed to change in response to the flow of spin-polarized currents. The spin-transfer torques exerted by vertical transport 
drive spin super-currents through the easy-plane nanomagnet.  
Below we first explain some key ideas by considering simpler macrospin limit in which the 
spin-stiffness in the thin film nano-magnet is strong enough to inhibit spatial 
variation in magnetization orientation, and then discuss the long magnetic link limit.    

\noindent
{\color{blue}{\em Macrospin Limit}}---
Because they are separated by non-magnetic spacer layers, the magnetizations 
of different magnetic circuit elements behave independently.  When the system is able to reach
a time independent state in the presence of bias voltages, the $\hat{z}$-direction 
spin of the easy-plane magnetic layer satisfies the torque-balance equation,
\begin{equation}
\hbar \frac{dS_{z}}{dt} = -2 K \sin(2\phi) + F_{L} g_{L} \mu_L +F_{R}  g_{R} \mu_R = 0, 
\label{static} 
\end{equation} 
where the first term on the right hand side is contributed by in-plane magnetic anisotropy, with $-K \cos(2 \phi)$ being the 
macrospin in-plane anisotropy energy. 
The remaining terms are spin-transfer torques~\cite{STT,spintransfertorquereview} applied at left and right electrodes by number currents 
\begin{equation}
I_{L,R} = g_{L,R} \mu_{L,R}/\hbar,
\label{eq:ivstatic}
\end{equation}
where $g_{L,R}$ is the stack conductance in $e^2/h$ units, and $\mu_{L,R}$ the circuit bias energy at 
the $L,R$ electrode. The currents contribute to $dS_{z}/dt$ 
because charge flows between electrodes with opposite 
perpendicular magnetization orientations; the factors $F_{L,R} \sim 1$ are 
material-dependent spin-injection efficiency factors,
and the conductances $g_{L,R}$ are relatively large because the in-plane nano magnet can 
efficiently reverse the spins of electrons which propagate through it.  Provided that
\begin{equation} 
\label{Ic1} 
|F_{L} I_{L}+ F_{R} I_{R} | <  I_{c} = 2 K/\hbar, 
\end{equation} 
Eq.~\eqref{static} has a solution
and the vertical conductance is local in the sense that $I_{L,R}$ depends only on 
$\mu_{L,R}$. Note that the static solution is allowed for any value of the individual currents 
as long as the total injected spin-current satisfies Eq.~\eqref{Ic1}. 
%As will be more apparent 
%when we consider long magnets to which the macro-spin model does not apply,
%large currents are possible because supercurrents carry spin between 
%electrodes without dissipation.  

When the total spin current exceeds $\sim 2K/\hbar$, Eq.~\eqref{static} can no longer be satisfied.
The magnetization can no longer reach a steady state value and begins to precess. 
Assuming that the easy-plane anisotropy is very large so that the magentization remains in the XY plane, Eq.~\eqref{static} then generalizes to

\begin{align}
\hbar \frac{dS_{z}}{dt} = &-2 K \sin(2\phi) - g_{i}\, \hbar \dot{\phi} \nonumber \\
& + F_L g_{L} (\mu_L - \hbar \dot{\phi}) + F_R g_{R} (\mu_R - \hbar \dot{\phi}) =0.  
\label{dynamic} 
\end{align} 

\noindent
Here the second equality keeps the spin direction close to the easy-plane and the three terms proportional to $\dot{\phi}$ 
capture magnetization decay due to Gilbert damping and spin-pumping~\cite{spinpumpingRMP,Caveat_Pumping} into the 
two electrodes.  The spin-pumping terms can be understood by performing a unitary transformation to a 
spin-direction frame which precesses with the magnetization.  The exchange field of the ferromagnet
is then static, allowing standard transport ideas to be applied, but the energy bias across the 
vertical transport stacks is shifted.  In the precessing case 
\begin{equation}
\hbar I_{L,R}  =  g_{L,R}  \; (\mu_{L,R} - \hbar \dot{\phi}).
\end{equation}
The coefficient $g_i$ in the Gilbert damping term is related to the usual dimensionless 
damping coefficient $\alpha$ by $\hbar g_{i} = \alpha S$ where $S$ is the total macrospin of the easy-plane 
magnetic nano particle.

\begin{figure}[!]
\includegraphics[width=3.3in]{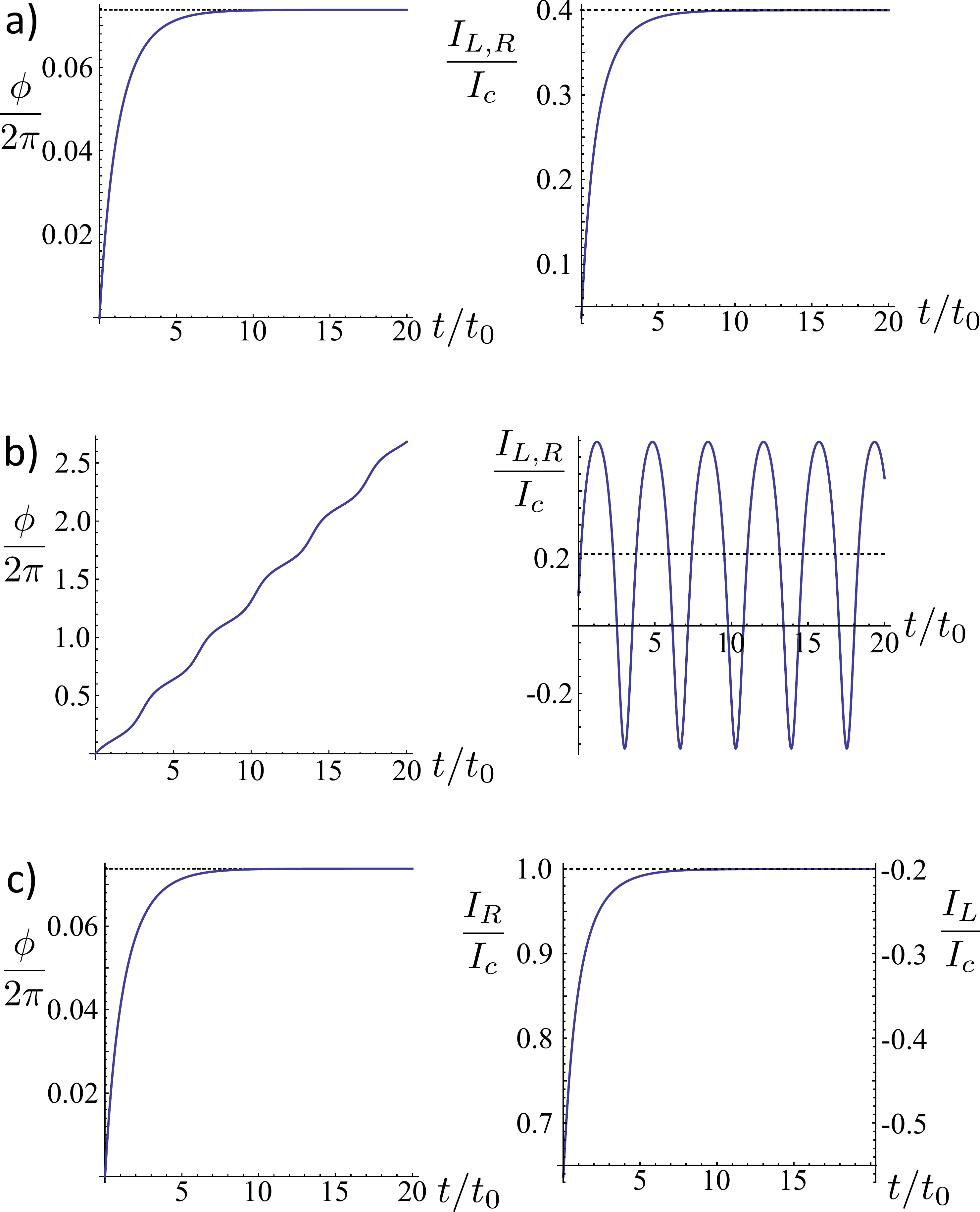}
\caption{\label{fig:two}(color online). 
Typical Magnetization Dynamics results for $g_{i}=g_{L}=g_{R}=g$, $F_{L,R}=1$ and initial value 
$\phi(t=0)=0$. The left panels depict the evolution of the macro-spin orientation $\phi$ and the right panels
illustrate the behavior of the currents in units of $2K/\hbar$.  In each case the dashed line shows the 
long-time average current.   a): $\mu_{L}=\mu_{R}= 0.8 K/g$ -- the total spin-current is below the critical value and the 
magnetization and currents approach time-independent values. b): $\mu_{L}=\mu_{R}=2.0 K/g$ -- the total spin current 
exceeds the critical value.  The magnetization precesses non-uniformly and the current has a large oscillating component.
The precession becomes more uniform and the oscillating currents weaker when the currents are increased 
further. c): $\mu_{L} = -0.4 K/g$, $\mu_{R} = 2.0 K/g$ -- 
the spin-current injected by the right contact is the same as in case b) but the total spin-current is the same as in case a). 
The time unit is $t_0=\hbar (g_i+g_L+g_R)/(4 K)$.}     
\end{figure}

Given $\mu_{L}$ and $\mu_{R}$, Eq.~\eqref{dynamic} can be solved for $\phi(t)$.
(Solutions for some typical parameters are presented in Fig.~\ref{fig:two}.) 
When the total current is comfortably in excess of the critical value, $\dot{\phi}(t)$ is approximately 
constant. In this case averaging over time yields the following non-local relationships between electrode
currents and biases: 

\begin{align} 
I_L  &= \frac{ g_L (g_{i}+F_R g_R)}{F_L g_L +F_R g_{R} + g_{i}} \; \mu_{L}    + \frac{- F_R g_R g_L}{F_L g_L +F_R g_{R} +g_{i}} \; \mu_R \nonumber \\
I_R &= \frac{ -F_L g_L g_R}{F_L g_L +F_R g_{R} +g_{i}} \; \mu_L +  \frac{g_R (g_{i}+F_L g_L)}{F_L g_L +F_R g_{R} +g_{i}} \; \mu_R
\label{eq:ivdynamic}
\end{align}

\begin{figure}[!]
\includegraphics[width=3.0in]{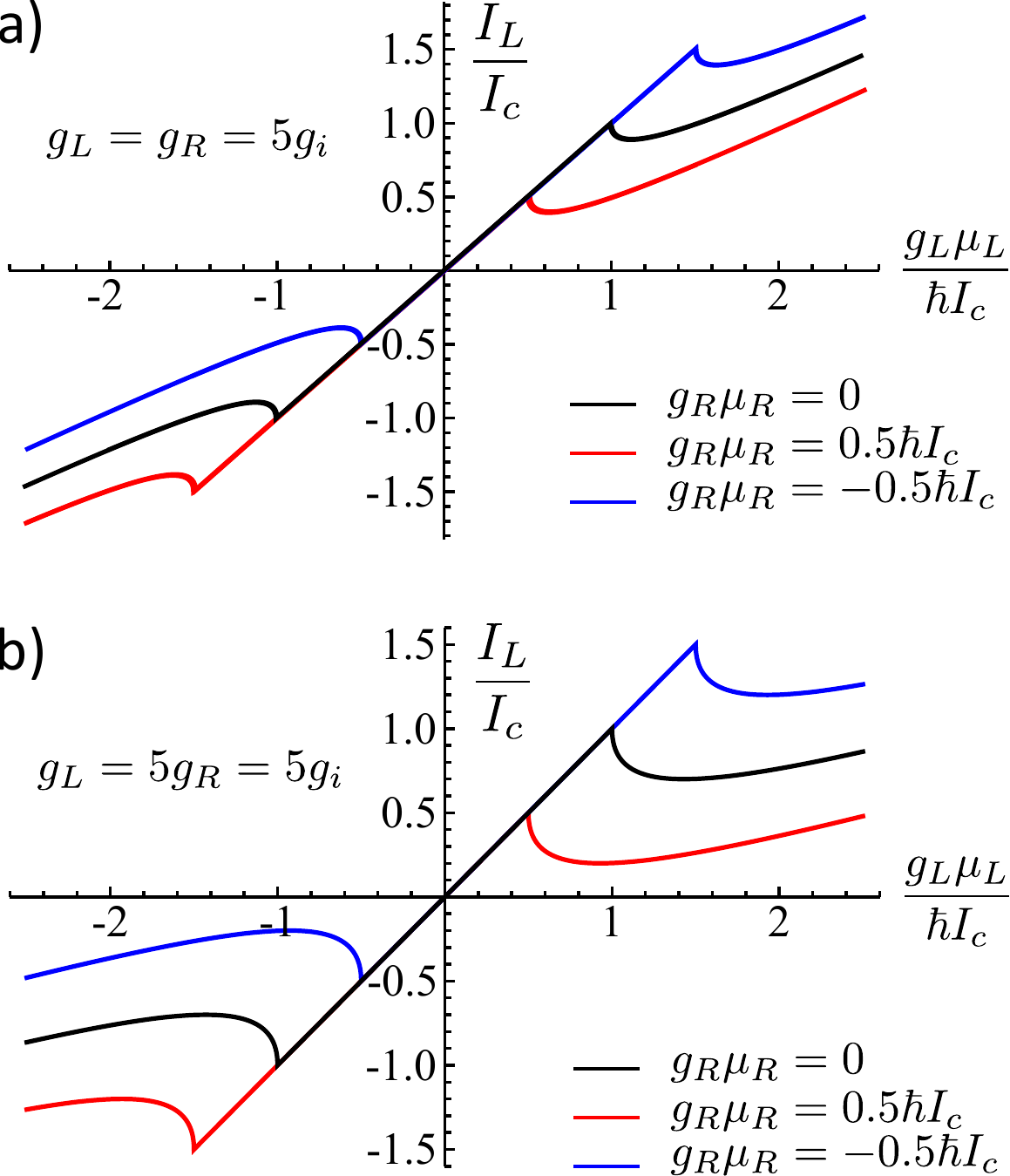}
\caption{\label{fig:three} (color online).
Current-voltage relationship of the left electrode for $\mu_{R}= 0$, $\mu_{R}=-K/g_R$ and $\mu_{R}=K/g_R$.  Currents in the dynamic regime are time averages over the limit cycle.  
Panel a) $g_L=g_R = 5 g_{i} $.  Panel b) $g_{L} = 5 g_{R}= 5 g_{i}$.  Note that 
the increase in resistance in the dynamic regime is larger when both $g_{R}$ and $g_{i}$ are much smaller than $g_{L}$.} 
\end{figure}

\noindent The crossover between Eqs.~\eqref{eq:ivstatic} which apply in the static magnetization regime
and Eqs.~\eqref{eq:ivdynamic} which apply far into the dynamic magnetization regime 
occurs rather abruptly as can be seen in Fig.~\ref{fig:three} 
where we plot the time-averaged current in the left electrode $I_L$ {\em vs.} $\mu_{L}$ for different values of $\mu_{R}$.   

Both the non-locality of transport and the contrast between the dynamic and static magnetization regimes are 
enhanced when spin-pumping is the dominant magnetization dissipation mechanism, {\em i.e.} when
$g_{L,R} > g_i$.  Achieving sufficiently large spin-injection into easy-plane 
nano magnets with Gilbert damping that is sufficiently small for this inequality to be satisfied
is a challenge which can now be met thanks to recent advances in 
spin-torque oscillator~\cite{SpinTorqueOscillator} technology.     
Because it has a relatively small interface resistance the Co/Cu materials combination is
favorable for the perpendicular-magnet/magnetic-spacer elements of the structure.
Similarly, because of its 
small Gilbert damping parameter, permalloy is an attractive material for the easy-plane anisotropy thin film. 
The Gilbert damping conductance $g_i$ of the permalloy nano magnet is proportional to its total spin and 
therefore to its volume.  The inequality we seek is favored by designing samples in which the easy-plane ferromagnet cross-sectional area 
is dominated by the portions within the metal stacks and not by the link portion.  In this limit $g_{L,R}$ and 
$g_i$ are both proportional to the stack cross-sectional area, while $g_i$ is in addition 
proportional to the permalloy thin film thickness.  Assumming a Gilbert damping parameter 
$\alpha \sim 10^{-2}$, and a spin per atom $\sim 2$, we conclude that $g_{L,R}$ can 
be larger than $g_i$ for films thinner than $\sim 4$ nm.  It is of course always possible to 
reduce $g_{L,R}$ below maximum values, for example by introducing a thin insulator layer in the metal stacks,
to reverse the sense of this inequality in one or both metal stacks. 
For 1 nm thick films and $\sim 10^{-14}$ m$^2$ stack cross-sections, we estimate that 
$g_{L,R} \le 300,000$ and, assuming that only $20 \%$ of the permalloy area is 
in the link segment, that $g_i \sim 80,000$.       
For a given sample shape, the in-plane magnetic anisotropy of thin film permalloy
can be controlled by varying growth conditions~\cite{Permalloy_Anisotropy}
to values as small as $\sim 10^{-8}$ eV per permalloy spin.  It follows that for
the same device dimensions, the bias energies necessary to drive the transition between static
and dynamic regimes can be as small as $\sim 10^{-7}$ eV.  
Electronic transport switches which operate at such small bias voltages are potentially attractive for
low-power-consumption electronic devices.    

\begin{figure}[!]
\includegraphics[scale=0.33]{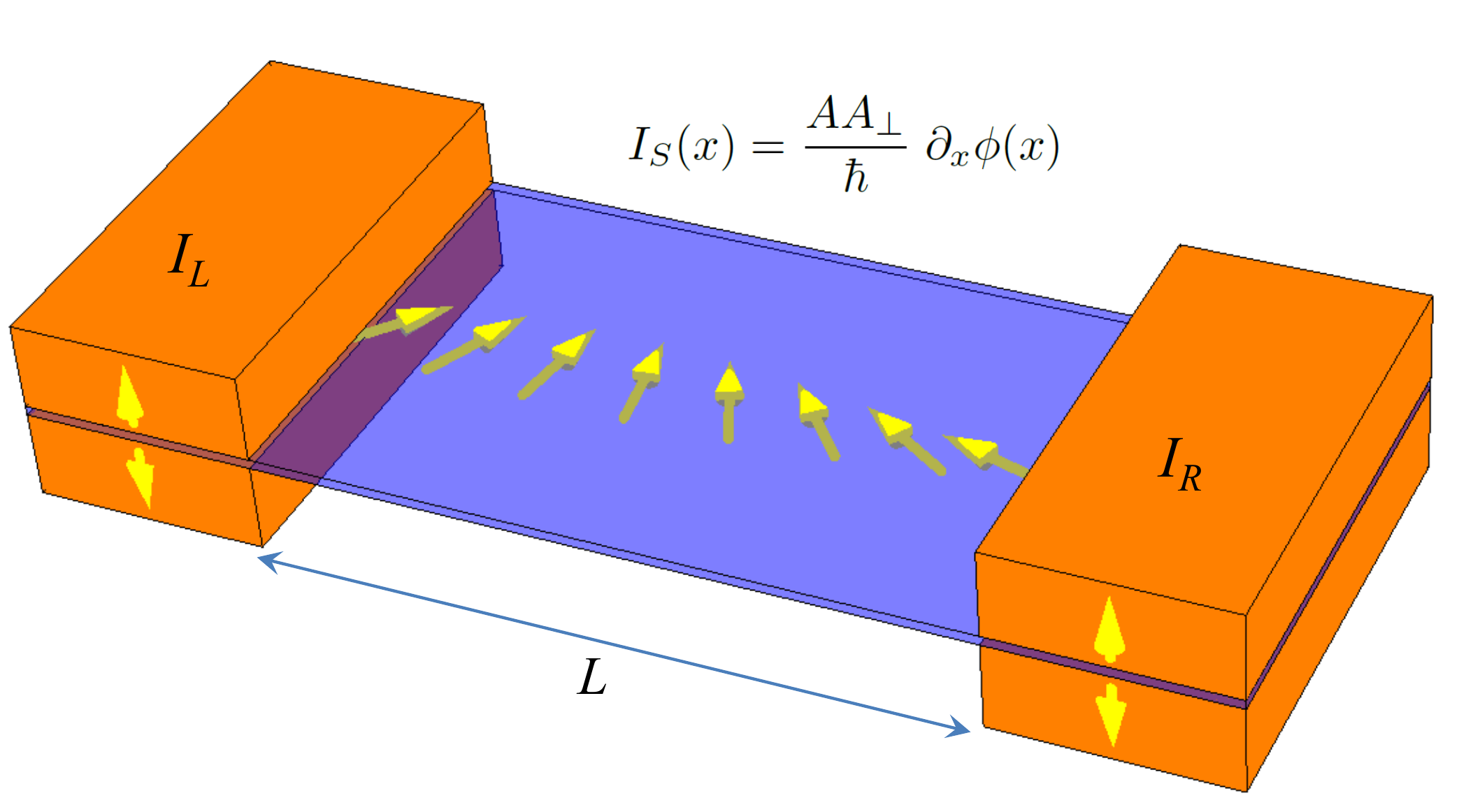}
\caption{\label{fig:four} 
(color online). Two vertical spin injectors coupled by an easy-plane ferromagnet.   The yellow arrows indicate 
magnetization orientation $\phi$ within the easy plane.  The spin-supercurrent is proportional to the spatial 
gradient of the magnetization orientation $\partial_x \phi$.}     
\end{figure}

\noindent
{\color{blue}{\it{Long Nanomagnet Limit}}}---
When the in-plane magnetic anisotropy is tuned to larger values, spatial variation of the magnetization 
orientation can become important.  In Fig.~\ref{fig:four} we illustrate the case of a 
long narrow permalloy thin film which provides a magnetic link between 
metal electrodes at left and right.   Assuming that the link is sufficiently narrow to force constant 
magnetization across the bar, the Landau-Liftshitz equation in this case is 
\begin{equation}
-2 \frac{K}{L} \sin(2\phi) - \frac{g_{i}}{L} \, \hbar \dot{\phi}+  A A_{\perp}  \partial_x^2 \phi  = 0,
\label{long} 
\end{equation} 
where $A_{\perp}$ is the cross-sectional area of the bar, $L$ is its length, and $A$ is the magnetic stiffness coefficient.
The two terms that are balanced in the static limit can be interpreted as contributions to the rate of change of the 
local $S_z$ density from local magnetization precession and from the divergence of the spin-supercurrent,
\begin{equation} 
 I_{S}(x) = \frac{A A_{\perp}}{\hbar}  \; \partial_x \phi(x).
\label{ISuper} 
\end{equation} 
Because the exchange splitting of typical ferromagnetic metals is much 
larger than the anisotropy energy per particle, the spin-current injected at left and right is
converted~\cite{Chen,Su} nearly locally into a spin-supercurrent.  It follows that 
\begin{eqnarray} 
I_{S}(x=0) &=&F_{L}  I_{L} =  F_{L} g_{L}  \; [\mu_{L} - \hbar \dot{\phi}(x=0)], \nonumber \\
I_{S}(x=L) &=& - F_{R} I_{R} = - F_{R}g_{R}  \; [\mu_{R} - \hbar \dot{\phi}(x=L)].  
\end{eqnarray}
As in the macrospin limit vertical transport at each electrode is local when the 
easy-plane magnetization orientation is static, but the switching boundary is dependent on 
both bias voltages.  Some insight into switching properties can be gained by noting that 
in the static case 
\begin{equation}\label{C}
\frac{\hbar^2 I_s^2}{2 A A_{\perp}} + \frac{K}{L} \cos(2\phi)  = C,
\end{equation} 
where $C$ is a constant independent of position.  
When $F_L I_{L}=-F_R I_{R}$ the spin-supercurrent has the same value at L and R 
bar ends, and static magnetization configurations are allowed at large 
current magnitudes.  When $I_{L}$ and $I_{R}$ have the same sign at the bar ends, however, 
$I_s$ must change sign as a function of position along the bar and static solutions are 
possible only when: 
\begin{equation} 
\max (F_L^2 I_L^2,F_R^2 I_R^2) < \frac{4 K A A_{\perp}}{L\hbar^2}.
\label{ic2}
\end{equation} 
For large $K$ this limit on the injected spin-currents is more stringent than 
Eq.~\ref{Ic1}.  In both cases the spin-injection limit depends on bias voltages in 
both metal stacks and the collective magnetic degree of freedom provides the 
coupling between electrodes. 

\begin{figure}[!]
\includegraphics[scale=0.6]{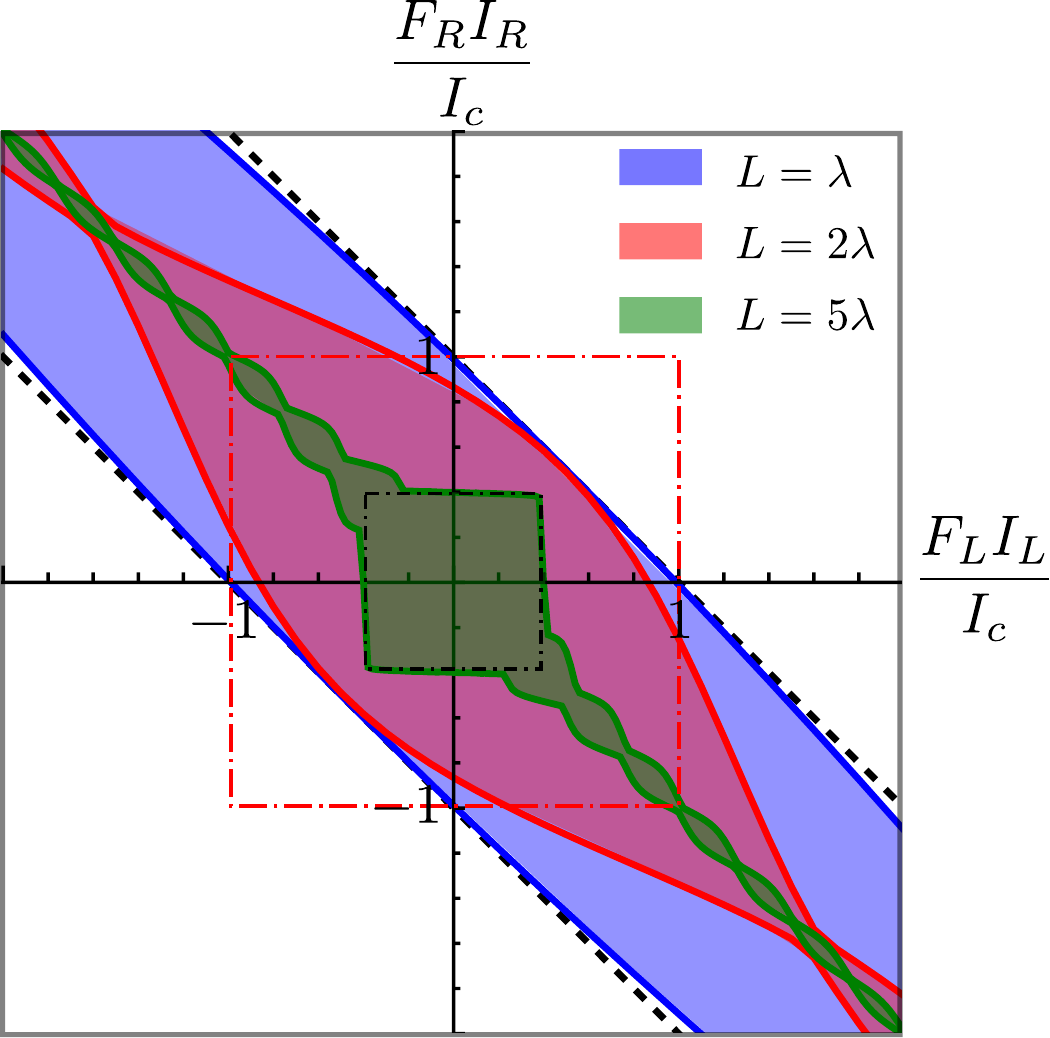}
\caption{\label{fig5} (color online).
Regions in the $I_{L,R}$ parameter space where static magnetization solutions exist. 
The blue, red, and green regions correspond to different ratios of the permalloy link length, $L$, to the length $\lambda$ defined in Eq.~\eqref{lambda}. The dashed black lines lines enclose the region predicted by Eq.~\eqref{Ic1} intended to apply in the macrospin model. 
The box enclosing the entire picture, the red dashed-dotted box and the black dashed-dotted box enclose the regions described by Eq.~\eqref{ic2},
intended to apply in the long-link limit, for $L=\lambda$, $L=2 \lambda$, $L=5 \lambda$ respectively. 
Note that the $L=\lambda$ case (blue region) is already well described by the macrospin model, whereas the $L=5 \lambda$ (green region) is already well described by Eq.~\eqref{ic2}, except for a narrow strip near $F_R I_R=-F_L I_L$.}     
\end{figure}

The ratio of the length of the permalloy link, $L$, to the typical magnetization variation length scale:

\begin{equation}\label{lambda}
\lambda\equiv\frac{1}{2}\sqrt{\frac{AA_\perp L}{K}},
\end{equation}

\noindent controls how far the system is from the macrospin limit. When $L \ll \lambda$ the critical currents are given by the macrospin result 
in Eq.~\eqref{Ic1}.
% (in addition the currents must be smaller than the critical value in Eq.~\eqref{ic2}).
In the opposite limit, $L \gg \lambda$, the currents at which steady solutions exist satisfy Eq.~\eqref{ic2} except 
within a narrow strip along the $F_R I_R=-F_L I_L$ line.  Figure~\ref{fig5} illustrates this behavior.
 
\noindent
{\color{blue}{\it{Discussion}}}---
The I-V characteristics discussed in this Letter provide an example of an electron transport
phenomenon in which a collective degree-of-freedom, the magnetization orientation, plays an active role. 
This type of phenomenon is of potential interest for electronic device applications mainly because 
it can lead to current flow response to bias voltage that is sharp on scales smaller than $k_{B} T/e$,
an impossibility for the single-particle transport processes exploited in most current electronic devices.
We have so far ignored the role of thermal fluctuations in the phenomena discussed here. 
Switching between static and dynamic magnetization configurations \cite{SunSwitch,RalphSwitch,AmbegaokarHalperin}
is stochastic with a thermal energy barrier due to anisotropy that vanishes at the critical currents or 
bias voltages.  It follows that reliable collective switching can be driven by 
changes in bias voltage $\delta V$ with $ e \delta V/k_{B} T \sim eV_{\rm switch}/K \sim 1/g_{L,R} \ll 1 $,
where $V_{\rm switch}$ is the bias voltage typically required for switching, 
exceeding the limit possible with switches based upon individual independent electron
behavior.

\noindent
{\color{blue}{\em Acknowledgements.}}---
HC, IS and AHM were supported by the Welch Foundation under Grant No.~TBF1473 and by the SWAN nano electronics program. ADK was supported by NSF-DMR-1309202 and the SRC-INDEX spin logic program.
 
 \bibliographystyle{apsrev4-1}

\begin{thebibliography}{100}
\bibitem{Caveat_CrossTalk}  The resistance of 1 nm thick permalloy thin films at room temperature is~\cite{Janak} 
$\sim 300$~$\rm{\Omega}$ per square.
The resistance area product of the Co/Cu metal stacks is typically near  
$10^{-15}$~${\rm \Omega}{\rm m}^2$~\cite{Bass}. 
For a permalloy link widths of $\sim 5 \times 10^{-8}$~m, 
and a lateral dimensions of the metal stacks of $\sim 10^{-7}$~m,
the link resistance ($\sim 500$~$\rm{\Omega}$) is substantially larger than the stack 
resistance ($\sim 0.1$~$\rm{\Omega}$).
Crosstalk has little influence on device performance unless the stack resistance 
is comparable to the link resistance.
\bibitem{Caveat_Pumping}  We have for simplicity assumed that the efficiency factors are identical for spin-transfer torques and 
spin-pumping.  This is strictly speaking true only when the source and drain electrodes are fully spin selective.
\end{thebibliography}

\end{document}